\documentclass[preprint2]{aastex}

\slugcomment{Accepted for publication in ApJ: March 26, 2007}

\shorttitle{ROTATION MEASURES BEHIND THE SOUTHERN GALACTIC PLANE}
\shortauthors{BROWN ET AL.}

\begin{document}

\title{Rotation Measures of Extragalactic Sources Behind the Southern Galactic
Plane:  New Insights into the Large-Scale Magnetic Field of the Inner Milky Way}

\author{J. C. Brown,\altaffilmark{1}  M. Haverkorn,\altaffilmark{2,3} 
B. M. Gaensler,\altaffilmark{4,5,6} A. R. Taylor,\altaffilmark{1}
N. S. Bizunok,\altaffilmark{5} \\
N. M. McClure-Griffiths,\altaffilmark{7} 
J. M. Dickey,\altaffilmark{8} and A. J. Green\altaffilmark{4}}
\altaffiltext{1}{Department of Physics and Astronomy, University of
Calgary, T2N 1N4, Canada; jocat@ras.ucalgary.ca; russ@ras.ucalgary.ca}
\altaffiltext{2}{Jansky Fellow, National Radio Astronomy Observatory}
\altaffiltext{3}{Astronomy Department, UC-Berkeley, 601 Campbell Hall, Berkeley, CA 94720; 
marijke@astro.berkeley.edu}
\altaffiltext{4}{School of Physics A29, The University of Sydney, NSW 2006, Australia; bgaensler@usyd.edu.au; agreen@physics.usyd.edu.au}
\altaffiltext{5}{Harvard-Smithsonian Center for Astrophysics, 60 Garden Street, Cambridge, MA 01238; 
nbizunok@cfa.harvard.edu}
\altaffiltext{6}{Alfred P. Sloan Research Fellow, Australian Research Council Federation Fellow}
\altaffiltext{7}{Australia Telescope National Facility, CSIRO, PO Box 76, Epping, NSW 1710, Australia; 
Naomi.McClure-Griffiths@csiro.au}
\altaffiltext{8}{Physics Department, University of Tasmania, Private Bag 21, Hobart TAS 7001, Australia;
John.Dickey@utas.edu.au}

\begin{abstract}

We present new Faraday rotation measures (RMs) for 148 extragalactic
radio sources behind the southern Galactic plane ($253\degr \le
\ell \le 356\degr$, $|b| \le 1.5\degr$), and use these data in combination
with published data to probe
the large-scale structure of the Milky Way's magnetic field. We
show that the magnitudes of these RMs oscillate with longitude in
a manner that correlates with the locations of the Galactic spiral
arms.  The observed pattern in RMs requires the presence of at least one
large-scale magnetic reversal in the fourth Galactic quadrant,
located between the Sagittarius-Carina and Scutum-Crux spiral arms.
To quantitatively compare our measurements to other recent studies,
we consider all available extragalactic and pulsar RMs in the region
we have surveyed, and jointly fit these data to simple models in
which the large-scale field follows the spiral arms.  In the
best-fitting model, the magnetic field in the fourth Galactic
quadrant is directed clockwise  in the Sagittarius-Carina spiral arm
(as viewed from the North Galactic
pole), but is oriented
counter-clockwise in the Scutum-Crux arm.  This contrasts with
recent analyses of pulsar RMs alone, in which the fourth-quadrant
field was presumed to be directed counter-clockwise in the
Sagittarius-Carina arm. Also in contrast to recent pulsar RM studies,
our joint modeling of pulsar and extragalactic RMs demonstrates
that large numbers of large-scale magnetic field reversals are not
required to account for observations.

\end{abstract}


\section{Introduction}\label{sec1}

The Galactic magnetic field is now recognized as a fundamental
component of the interstellar medium, and plays a critical role in
the formation and evolution of structures in the Milky Way.  An
important prediction in models of the large-scale magnetic field,
in both the Milky Way and in other galaxies, is the existence of
magnetic field reversals (regions of magnetic shear across which
the field changes direction by roughly 180$^\circ$).  Determining
the number and location of these magnetic reversals is essential
to understanding Galactic evolution \citep{anvar05}.  While the
majority of  the recent studies suggest several large-scale
reversals in the Galaxy along its radius,  it is interesting to
note that most other galaxies exhibit either one reversal or none
\citep{beck06}.   Is our Galaxy unique this way, or is it simply a
difference in observing methods?

The large-scale Galactic magnetic field is concentrated in the disk,
and is most often studied via observations of rotation measure (RM),
the measurable consequence of Faraday rotation.  For a source that
emits linearly polarised radiation at an angle $\phi_\circ$,  the received
radiation will have a polarisation angle at 
a wavelength, $\lambda$ [m] ,  given by:
\begin{equation}
\phi  =  \phi_\circ + \lambda^2 \; 0.812 \int{n_e}{\bf B}\cdot {\bf
dl} =   \phi_\circ + \lambda^2 \;  {\rm RM} \\
\end{equation}
where $n_e$ [cm$^{-3}$] is the electron density and {\bf B} [$\mu$G]
is the magnetic field along the propagation path {\bf dl} [pc].
The RM integral  is over the path from the source  to the observer.

\begin{figure*}
\epsscale{2}
\plotone{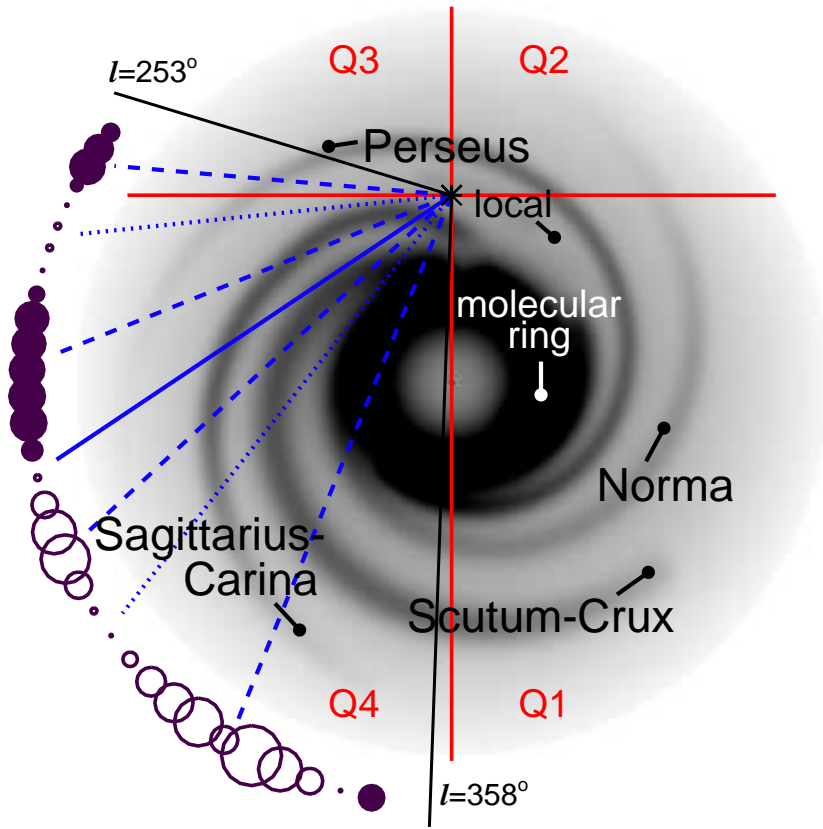}
\caption{View of the Galaxy from above. The grey scale is the CL02 electron density model, with labels indicating the spiral arms, and
the asterisk indicating the location of the Sun.  Quadrants 1, 2, 3, and 4 are labeled with Q1, Q2,
Q3 and Q4, respectively.  The bounding longitudes of the SGPS (Phase I) are indicated by black lines and are labeled. The circles represent the smoothed-averaged SGPS RMs as shown in the middle
panel of Figure {\ref{fig3}}.  Filled (open) circles indicate positive (negative) RM, with the size
of the circles linearly proportional to $|$RM$|$,  truncated between 59 and 592 rad m$^{-2}$
(the $|$RM$|_{max}$ from the middle panel of Figure {\ref{fig3}}) so that sources with
$|$RM$| < 59$ rad m$^{-2}$ are set to 59 rad m$^{-2}$. 
The blue dashed (dotted) lines are also from the middle panel of 
Figure \ref{fig3}, and indicate the approximate longitudes of $|$RM$|$ maxima (minima) in the 
SGPS RM data.  The solid blue
line is the longitude where the RMs transition from primarily positive to primarily
negative ($\ell \sim 304\degr$).}
\label{fig1}
\end{figure*}

For more than four decades, the RMs of both pulsars and extragalactic
radio sources (EGS) have been used to probe the Galactic magnetic
field. This has led to a series of clear conclusions.  First, in
the local spiral arm, the field is unquestionably directed clockwise
(CW), as viewed from the North Galactic pole \citep{man72, man74,
heiles96a}.  The field in the  first quadrant (Q1; $ 0\degr\le\ell\le90\degr$; see
Figure \ref{fig1}) of the
Sagittarius-Carina spiral arm is reversed relative to the local arm
\citep{TN80, sk80, ls89}, implying a large-scale field reversal
between the local arm and the Q1 component of the Sagittarius-Carina
arm.  Some evidence suggests that this reversed field extends into
the fourth-quadrant (Q4; $270\degr\le\ell\le360\degr$) component of  the Sagittarius-Carina arm
\citep[eg.][]{rl94,han99,frick01}. 

At larger distances, the existence of other large-scale reversals
in the Galaxy remains unclear.  Ideally, reconstruction of the
Galactic magnetic field should utilize information from both pulsars
and extragalactic sources.  However, until recently there have been
very few EGS RMs available at low-latitude that
might be used to study the field in the disk.  Studies that have
utilized pulsar RMs alone are constrained by the comparatively
sparse sampling of pulsars on the sky, which can make it difficult
to map the field in complicated regions.

Three recent pulsar RM studies are as follows.
Using pulsar RMs, \citet{wc04} investigated the existence
and location of reversals in Q1, concluding that a reversal occurs
between each arm so that the magnetic fields in adjacent arms are
oppositely directed.  While the  evidence presented by \citet{wc04} 
for a reversal between
the local arm and the Sagittarius-Carina arm  in Q1 is indisputable,
their evidence for additional reversals is based on  limited data,
and they acknowledge that the evidence for a reversal into the
Q4-component of the Sagittarius-Carina arm is not well-defined.

\citet{vallee05} investigated azimuthal field configurations in the
Galaxy using pulsar RMs averaged in concentric rings.  He concluded
that best-fit for this model was an overall clockwise magnetic field
with a 2 kpc wide counter-clockwise (CCW) ring, located between 4
and 6 kpc from the Galactic center (note that a non-standard 
Solar-circle radius of 7.2 kpc was assumed in this study).  In this model, the 
Galactic field has two
large-scale reversals, and the Q1 component of the Sagittarius-Carina
arm is CCW, while the Q4 component is CW.

Using 223 new pulsar RM observations primarily in the fourth quadrant,
in conjunction with previous pulsar RMs, \citet{han06} concluded
there is a reversal at every arm-interarm boundary, so that the
fields in the arms are  directed CCW, and the interarm regions are
directed CW.  A potential inconsistency with this model is that  the
majority of the pulsars distributed along the Q4 component of the
Sagittarius-Carina spiral arm  have what \citet{han06} describe as
`unexpectedly positive' RMs, which they suggest is the influence
of HII regions along the  lines-of-sight to the affected pulsars
\citep[see also][]{dm03}.

Recent surveys  of the Galactic plane at high resolution and at
multiple wavelengths have assisted greatly in the study of the Galactic
magnetic field by addressing the previous paucity of low-latitude
EGS RM data \citep[eg.][]{bt01}.   One such survey is the Southern Galactic Plane Survey
\citep[SGPS;][]{mdg05,hgm06}.  The SGPS images low latitudes in the
third and fourth quadrants of the Galaxy, complementing the Canadian
Galactic Plane Survey in the northern hemisphere \citep[CGPS;][]{cgps}.
Rotation measures calculated from these data were used by \citet{mh06a}
to explore the structure of the small-scale field.  Here, we present
the tabulation of these RMs and use them to examine the structure
of the large-scale field.

\section{Rotation Measure Calculations}\label{sec2}

The  initial set of observations for the SGPS (ie. Phase I) spans an area of $253\degr \le \ell \le 358\degr$ and $|b| \le 1.5\degr$.
The observations were done with the Australia Telescope Compact Array (ATCA) in New South Wales, Australia.  For details about the SGPS observations and polarimetric data reduction,  see
\citet{bmg01} and \citet{hgm06}. 

RMs were calculated using  twelve separate 8 MHz bands 
centered on frequencies between 1336 MHz and 1432 MHz.
 The proximity of the 8 MHz bands allows for unambiguous RM calculations 
using the algorithm designed by \citet{btj03} for the CGPS, with appropriate modifications for the ATCA.  Specifically, an RM calculation was considered reliable if the source was sufficiently polarized
 ($>$0.3\%),
had sufficient signal to noise ($>$2) across at least 3 pixels, was Faraday thin \citep{vallee80}, and had a consistent value across the source.

Of the 215 polarised source candidates identified, 
148 sources had RMs that successfully passed the screening tests discussed above. These new 
data are given in Table 1. Three
of our 148 sources had a RM determined by \citet{bmg01}, as part of the SGPS test region
($325.5\degr \le \ell \le 332.5\degr, -0.5 \le b \le 3.5\degr$).   The previously determined values of these
sources are within the errors of the new values quoted in Table 1. 
These sources are indicated by footnote
marker $d$.  Two additional test region RM sources fall within the latitude range of the SGPS but they
lie within the noise perimeter 
of our data \citep[see][]{bmg01} and were not observable. There is one other previously
observed RM from an EGS, reported by \citet{broten88}, that falls within our field  at  $\ell=307.1\degr$, $b=1.2\degr$. This source is resolved in the SGPS and exhibits
a large gradient across the source. Consequently, it failed the screening, though its calculated
RM ($+183 \pm 23$ rad m$^{-2}$) is in agreement with the previously determined value ($+185 \pm 1$ rad m$^{-2}$).

\section{Observations of the Galactic Magnetic Field} \label{sec3}

\begin{figure*}
\epsscale{2}
\plotone{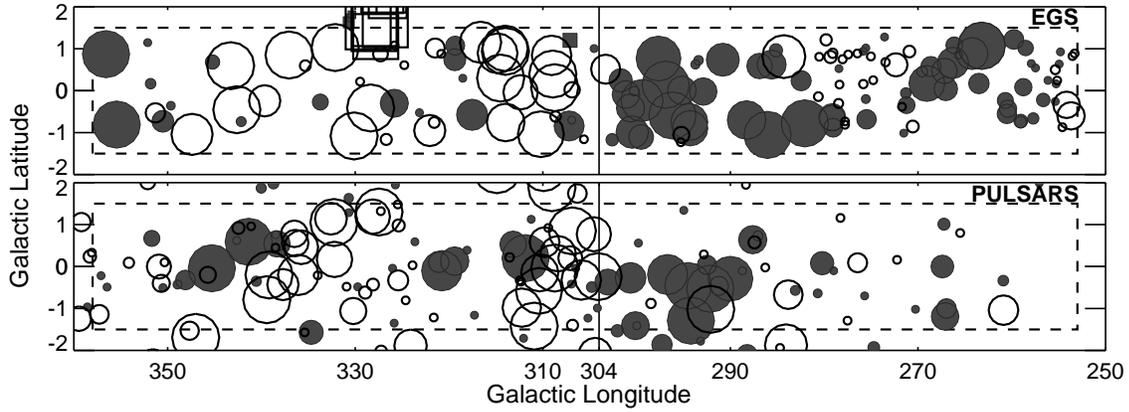}
\caption{Rotation measures for sources in the SGPS. Top panel: extragalactic RM sources; Bottom panel: pulsar RM sources.  Grey filled symbols indicate positive rotation measure, black open symbols indicate negative rotation measures; sizes of symbols are linearly proportional to the magnitude of RM truncated between 100-600 rad m$^{-2}$, so that sources with $|$RM$| < 100$ rad m$^{-2}$ are set 
to 100 rad m$^{-2}$, and those with $|$RM$| > 600$ rad m$^{-2}$ are set to 600 rad m$^{-2}$. 
In the top panel, the square at $\ell= 307.1\degr$, {\it b}$= 1.2\degr$ represents the EGS RM
 previously observed by \protect\citet{broten88}, while the squares at 
$\ell \sim 330\degr$ represent the EGS RMs of the SGPS test
region. The circles in the top panel represent the SGPS EGS as given in Table \ref{tab1}. 
The vertical line at $\ell= 304\degr$ is the  approximate longitude where the EGS RMs change sign. The dashed boxes indicate
boundaries of the SGPS region.}
\label{fig2}
\end{figure*}

Figure \ref{fig2} shows the RMs  of  EGS and pulsars in the SGPS region.  Most of the pulsars (99 of 120)
 are from \citet{han06}. The remaining are from \citet{pulsars} and \citet{han99}.  The most striking 
feature of Figure \ref{fig2} is the  change in sign
of RM from predominantly positive at low longitudes to predominantly negative at high longitudes
for both the pulsars and EGS at $\ell\sim 304^\circ $, though it is more prominent for the EGS.   
A change in sign of the overall trend in RM can only
come from a change of direction in the dominant line-of-sight component of the magnetic field.  The fact
that the RM sign change is abrupt indicates that this directional change is the result of a
physical field reversal, rather than a viewing angle effect such as that observed towards
$\ell \sim 180\degr$ \citep[eg.][]{bt01}. Furthermore, the abruptness indicates a thin current sheet and an
associated large gradient in
the  magnetic field \citep{heiles96}.

\begin{figure*}
\epsscale{2}
\plotone{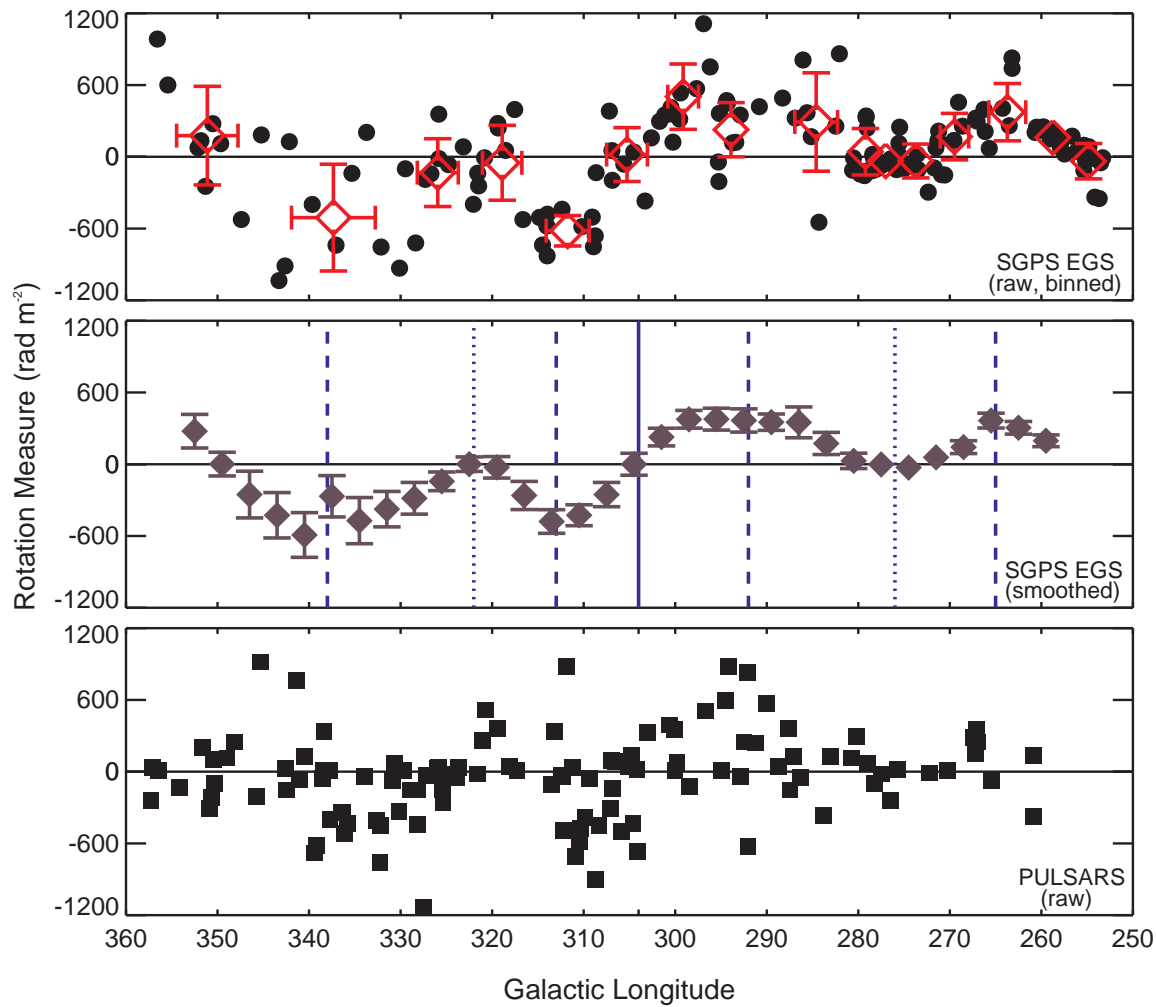}
\caption{RM versus Galactic longitude for RM sources in the SGPS region.  
Top panel: circles represent the individual SGPS EGS
sources (errors are smaller than the symbol size), while open red diamonds represent data 
averaged into
into independent longitude bins containing 9 sources (the end bin at 255$\degr$ contains 13 sources).  Where symbol size permits, the error bars are the standard deviation in longitude and RM for each bin.
Middle panel:  purple diamonds represent boxcar-averaged SGPS EGS data over 9 degrees in longitude with a stepsize of 3 degrees. In contrast to the binned data in the top panel, the error bars are the standard error of the mean.  The solid line marks the approximate longitude of transition from predominantly positive RMs to negative RMs  ({\it l} $\sim 304\degr$)  Dotted lines (dashed lines) mark approximate longitudes of minimum  (maximum) $|$RM$|$ in SGPS data. 
Bottom panel: squares represent the individual pulsars with known RM in the SGPS region (errors are 
smaller than the symbol sizes). }
\label{fig3}
\end{figure*}

By averaging the  EGS RMs in $\ell$ across the sky to reduce small-scale variations (for example, due to the small-scale field or 
intrinsic effects from the  EGS themselves), we can obtain more information about the large-scale field. 
The top panel in Figure \ref{fig3} shows the RM data from the SGPS, both raw and binned plotted
as a function of Galactic longitude. The 
middle panel shows these data smoothed. 
In all three presentations of the EGS data, an oscillating pattern of RM with longitude is visible.  The
transition from positive to negative RMs remains at $\ell \sim 304\degr$, indicated by the  solid vertical line.  In the bottom panel of Figure \ref{fig3}, we plot the individual pulsar RMs as a function of Galactic longitude. These data were not averaged in $\ell$ because of the variation and  uncertainty in the pulsar distances.   In spite of this, there are features seen here similar to the oscillatory pattern observed
in the EGS data.  

The dashed and dotted lines in the middle panel of Figure \ref{fig3} are the approximate longitudes 
of $|$RM$|$ maxima and minima respectively for the SGPS data. In Figure \ref{fig1}, we show
 these lines as 
lines-of-sight overlaid on
a top-down view of the Galaxy where the grey scale is the Galactic electron density model of \citet{cl02}, hereafter CL02.    Interestingly, the blue dashed lines ($|$RM$|$ maxima) tend to have the longest continuous fraction of their length along a spiral arm  or through the central annulus put in the CL02 to  correspond  to 
 the molecular ring.   This is consistent with the
expectation that EGS RMs should be dominated by the spiral arms as a result of the higher electron densities in these regions \citep{han06}. 
The strong positive RMs
 in the longitude range around $\ell \sim 292\degr$, seen in both the EGS and pulsars,  suggest the magnetic field in the Q4-component   of the Sagittarius-Carina arm is
 directed CW. The same conclusion was reached by \citet{caswell04} using data from a distant
 supernova remnant.   This CW field presents a simple explanation for the  positive pulsar RMs 
 identified by \citet{han06} as part of a  `Carina anomaly'. 
 
 Similarly, the strong negative RMs in the longitude range around $\ell \sim
 312\degr$ suggest the magnetic field in the Q4-component  of the Scutum--Crux arm is directed CCW.   Therefore, a magnetic field reversal must reside between 
the Sagittarius--Carina arm and the Scutum--Crux arm in Q4.  
Furthermore,  the strong evidence for
 a field reversal between the local and Sagittarius--Carina arms in Q1 suggests the reversal must slice through the Sagittarius--Carina arm if this reversal is 
 continuous between Q1 and Q4, such as in the model proposed by \citet{vallee05}.  
 The subsequently implied lack of alignment of the magnetic field with the Sagittarius-Carina
 arm is consistent with earlier observations of  at least a $5\degr$
offset in pitch angle between the orientation of the local field and that of the local spiral arm \citep{beck06}.

\section{Modeling the Large-Scale Magnetic Field}

As a separate approach to qualitative observational analysis of the data, we can 
 fit global magnetic field models to the RM data  (with assumptions
regarding  the electron distribution in the Galaxy), to explore the structure of the uniform field and
the location or existence of magnetic field reversals. 
 Here we use the CL02 electron density model and the technique of \citet{brown02} which
 uses linear inversion theory \citep{menke} to obtain a least-squares fit to the data. 
With this method, the boundaries of  magnetic field regions are fixed {\it a priori} but the strength 
and direction of the field within these regions are not.  
The model is fit to the observed RMs to derive the strength and direction within each region. 
This contrasts previous analyses 
 which also employed separate models for the electron density and magnetic field, 
but where the direction of the field was an input to the model \citep[e.g.][]{id98}. 

\begin{figure*}
\epsscale{2}
\plotone{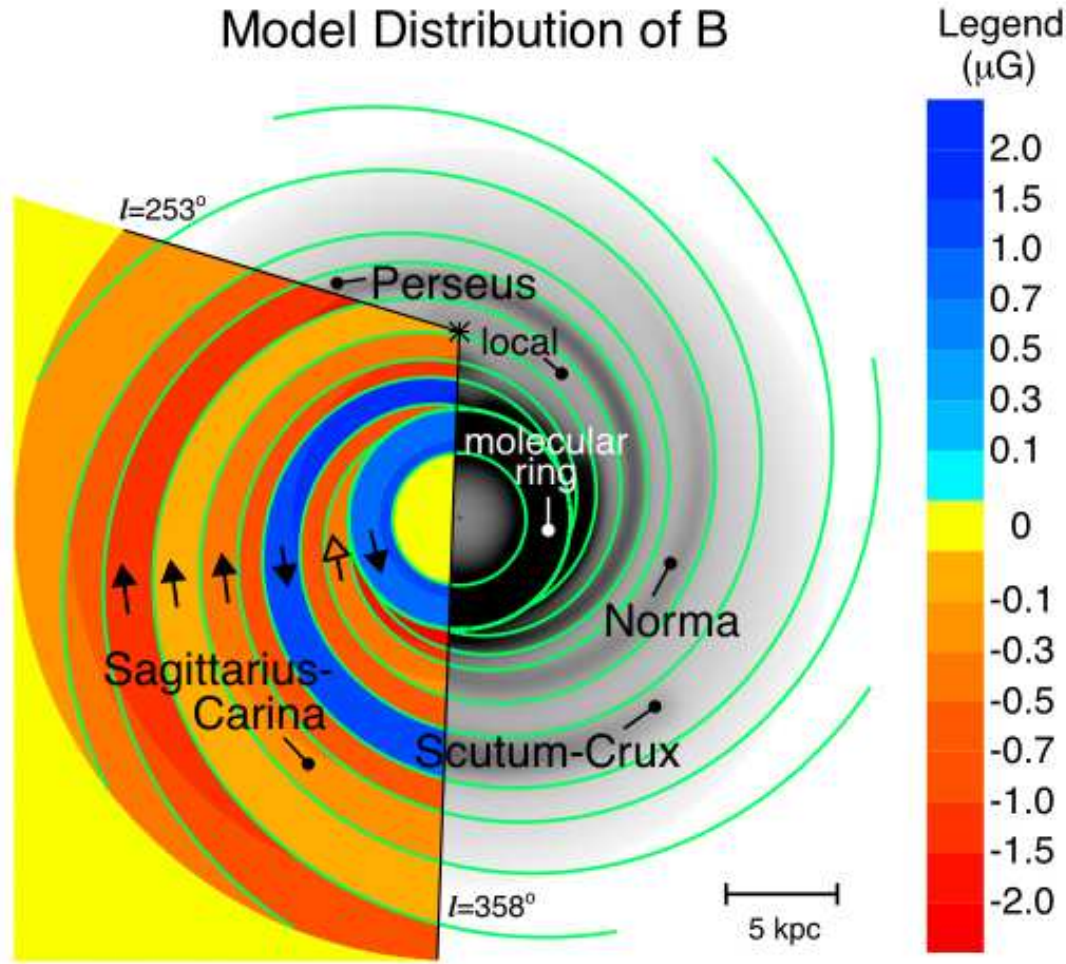}
\caption{A simple model of the magnetic field in the SGPS region, constrained using individual
RMs of 149 EGS and 120 pulsars. 
Colored regions indicate the total strength for the model fit of magnetic field corresponding to the 
regions identified by the  green lines as
discussed in the text. Red (blue) shading indicates a clockwise (counter-clockwise) field direction as viewed from the Galactic north pole. The direction of the field within the arms is also indicated by
arrows.  The open-head arrow on the Norma arm indicates the field direction in this arm is 
not well-constrained in this model configuration with the data used.  The bounding longitudes of the SGPS (Phase I) are indicated by black lines and are labeled. The grey scale is the CL02 electron
density model with labels indicating the spiral arms. 
}
\label{fig4}
\end{figure*}

We present here a simple model where we
consider nine magnetic field regions, eight of which are  either arm or interarm
regions, and the ninth corresponds to the molecular ring.  This model does not include the differing pitch angle between the magnetic and 
spiral arms discussed in section \ref{sec3}, but is instead designed to be directly comparable with
the results  of 
\citet{wc04} and \citet{han06}.  The region boundaries
are delineated by the green spirals in Figure \ref{fig4}. The constraints for the model
field within each arm or interarm region are:   1) a log-spiral with a pitch angle of 11.5$\degr$ as shown; 2)  a field strength that is inversely proportional to Galactic radius \citep[eg.][]{heiles96, beck01, btwm03}; 3)  a zero  vertical component; 4) 
a coherent direction within each region.  We set the magnetic field within the  molecular ring to have the same constraints as in the arms, except that the field is assumed to be purely azimuthal. 
At Galactic radii  less than 3 kpc  or
 greater than 20 kpc, or at more than 1 kpc from the mid-plane, the
 field is set to zero.  
 
We constrain the model using RM data from individual sources only within the SGPS region 
(120 pulsars, 148 SGPS EGS, and 1 EGS from Broten et. al 1988; see sections \ref{sec2} and \ref{sec3}).{\footnote{Global models using data from all quadrants is  beyond the scope of this paper, but will be presented in a future paper.}}  
For modeling purposes, we assume that the intergalactic
 contribution to the  EGS RMs is negligible \citep{sk80, gaensler05}, so that the EGS may be considered
 to reside at Galactocentric radii of 20 kpc.   Most of the published pulsar distances used are
 based on the CL02 model, and are therefore consistent with this model.
As a consequence of the limited sky coverage of the RM data used here,  we confine our analysis of the result to within  the SGPS longitudes. 
 
\begin{figure*}
\epsscale{2}
\plotone{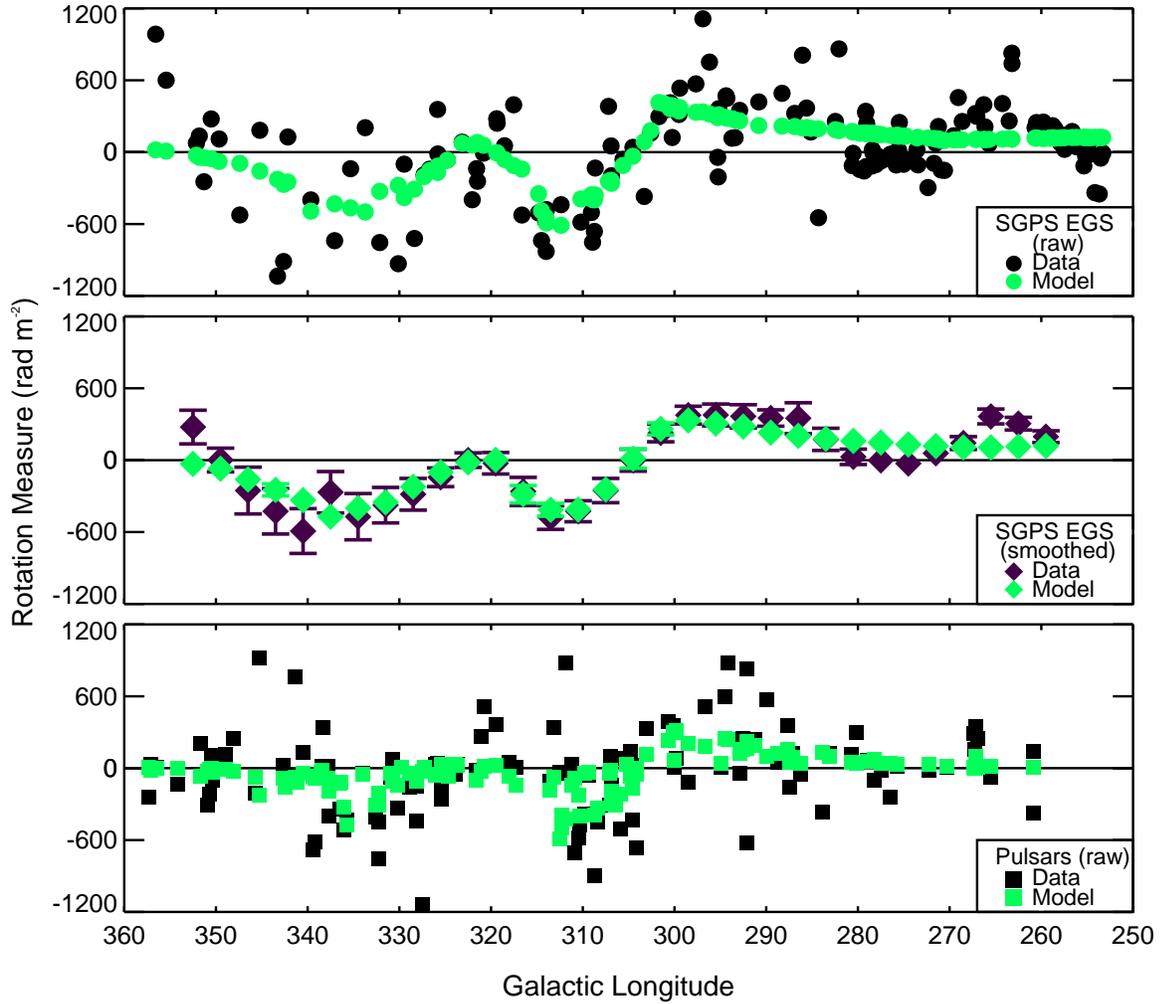}
\caption{Comparison of  modeled RMs (green symbols) and observed RMs. The modeled RMs are
calculated using the CL02 electron density model and the magnetic field model shown in Figure \ref{fig4}.  
The format of this figure follows that of
Figure \ref{fig3}, where the individual SGPS EGS are shown in the top panel, smoothed-averaged 
SGPS EGS data are shown in the middle panel, and the individual pulsars are presented in the
bottom panel. }
\label{fig5}
\end{figure*}

The best-fit output for this model is a CW field everywhere except for a CCW field in the Scutum-Crux arm and in the molecular ring,  as shown in Figure \ref{fig4}. 
Figure \ref{fig5} shows a plot of  both the measured and
 modeled RMs  for the individual SGPS EGS data (top panel), these data averaged and smoothed 
 as in the middle panel of Figure \ref{fig3} (middle panel), and the  measured and
modeled RMs for individual pulsars (bottom panel). 

This model is able to closely reproduce the RM
structure seen in the smoothed SGPS EGS data (recall that the fit is to the individual EGS and pulsar data) .  However, there are two  (relatively small)  discrepancies between the model and the observed data towards the outer Galaxy.  The first is in the 
vicinity around  $\ell \sim 275\deg$, where the measured data are more negative than the 
modeled data as also seen in the top panel Figure \ref{fig5}.  In Figure \ref{fig2}, there is a contained region of small, negative RM   
between $270\degr <l<283\degr$ at $b > 0\degr$, resembling a magnetic bubble  like that discussed by \citet{clegg92} and \citet{bt01}.  RMs at these longitudes nominally should be dominated 
by the field in
the local arm.  Thus, if the negative RMs seen here were to be attributed to the large-scale field, this would imply that  the field is directed counter-clockwise in the local arm. This is contrary to the many studies  that show the field is clockwise in the local arm, as discussed in section \ref{sec1}.  Interestingly,   
this region was previously identified in the Parkes 2.4-GHz survey as containing a polarized feature of unknown origin \citep{duncan97}.   Since the lines-of-sight in the outer Galaxy are considerably
smaller through the interstellar medium, compared to lines-of-sight through the inner Galaxy,
it is likely that this localized feature is dominating the RM for these EGS.  As a result,
averaging the negative RMs through this localized feature with the otherwise positive RMs 
of the local arm 
creates the effect of  ${\overline{\rm RM}} \sim 0$ at $l \sim 275\degr$.    

The other discrepancy is around $\ell \sim 265\degr$ where the measured RMs are 
more positive than the modeled RMs.  Lines-of-sight at this longitude are again through the outer
Galaxy where localized features including the Gum nebula  \citep{cs83} could dominate the RM.  
As seen in the top panel of Figure \ref{fig3}  there are  two EGS sources at $\ell = 263\degr$ that have RMs roughly double that of neighboring sources which biases the average RM 
near this longitude. Interestingly, there is also a localized peak in the pulsar data near this longitude, 
as seen in the bottom panel of Figure {\ref{fig3}}.   
  The model tries to fit both the negative region around
$\ell \sim 275\degr$ and the more positive region around $\ell \sim 265\degr$, with the
result being a compromise between the two.

Although the individual pulsar data are noisier than the averaged EGS data,  the trends of the pulsar 
data are also reproduced  by this model.
In particular, the model supports our observational conclusion that the Q4-component of the Sagittarius-Carina arm is directed CW, while the Q4-component of the Scutum-Crux arm is directed CCW.

Finally, we note that the direction of the Norma arm field  is not well 
constrained by this model or by the data.
Regardless of whether the field in the Norma arm is oriented CW or CCW, the results from this model contrast
the previous suggestions of reversals with every arm \citep{wc04} or at every 
arm-interarm boundary \citep{han06}.

 \section{Summary}
 
We present the rotation measures for 148 extragalactic sources found in the southern Galactic
plane survey.  The oscillations of rotation measure with longitude revealed by these sources, and 
as also seen in pulsar RM data, 
highlight the dominating effect of the spiral arms on rotation measure. Both empirically and with
a direct fit to measurements, the new data
show conclusively that the field is directed clockwise in the fourth-quadrant component of the
Sagittarius-Carina arm, and that a field
reversal exists between the Sagittarius-Carina arm and the Scutum-Crux arms in the fourth quadrant. 

A definitive measurement of the number of large-scale magnetic reversals in the Galaxy can only emerge from an analysis that includes pulsar and EGS RM data at all Galactic longitudes, and which considers a wide range of distinct field configurations.  In addition, the technique presented here is 
constrained to geometries imposed by the CL02 model.  With these caveats, the results from
our study of southern RMs  indicate that far fewer magnetic reversals are needed to explain 
the data than other recent studies have suggested.

\clearpage

\acknowledgments

We thank the anonymous referee for the insightful comments that have improved this manuscript. 
The Australia Telescope is funded by the Commonwealth of Australia for operation as a National
Facility managed by CSIRO. 
This work was facilitated in part by an associateship grant to JCB by the Alberta Ingenuity Fund.
MH acknowledges support from the
National Radio Astronomy Observatory, which is operated by Associated
Universities, Inc., under cooperative agreement with the NSF. 
 BMG  and MH acknowledge the support of the NSF  through grant
AST-0307358.


\clearpage

\begin{deluxetable}{rrllrrr}
 \tablewidth{321.2pt}
 \tablecaption{Rotation Measures of the SGPS }
 \tablehead{
\colhead{$\ell^{\;a}$}        & \colhead{$b^{\;a}$}      &
 \colhead{$\alpha$}   & \colhead{$\delta$}  &
 \colhead{$I^{\;b}$}        & \colhead{$m^{\;c}$}    &
 \colhead{ RM } \\
 $(^\circ)$\hspace*{2.2mm} & $(^\circ)$ \hspace{1mm} &
 \hspace{1mm} (J2000) & \hspace{1mm} (J2000) &
 (mJy) & (\%) &
 (rad m$^{-2}$)
}
 \startdata
253.30 &      0.89 & 08 19 46.2 & -34 42 05 &
179 &       5.8 & $          -8$ $\pm$ 12 \hspace{0.5mm} \\
253.52 &      0.83 & 08 20 08.4 & -34 55 20 &
21 &       6.4 & $         -50$ $\pm$ 26 \hspace{0.5mm} \\
253.68 &     -0.60 & 08 14 47.2 & -35 51 08 &
72 &       3.5 & $        -349$ $\pm$ 27 \hspace{0.5mm} \\
254.16 &     -0.34 & 08 17 10.3 & -36 06 01 &
64 &       5.5 & $        -338$ $\pm$ 19 \hspace{0.5mm} \\
254.60 &     -0.87 & 08 16 12.7 & -36 46 05 &
62 &       4.3 & $         -15$ $\pm$ 24 \hspace{0.5mm} \\
254.81 &      0.93 & 08 24 08.9 & -35 55 21 &
87 &       13.7 & $+84$ $\pm$ 13 \hspace{0.5mm} \\
254.95 &      0.63 & 08 23 20.7 & -36 11 56 &
41 &       4.9 & $+8$ $\pm$ 32 \hspace{0.5mm} \\
255.16 &      0.24 & 08 22 21.9 & -36 36 03 &
92 &       5.3 & $         -35$ $\pm$ 16 \hspace{0.5mm} \\
255.27 &      0.16 & 08 22 20.2 & -36 43 53 &
84 &       3.4 & $+26$ $\pm$ 23 \hspace{0.5mm} \\
255.36 &     -0.26 & 08 20 52.1 & -37 02 48 &
136 &       9.0 & $+97$ $\pm$ 10 \hspace{0.5mm} \\
255.36 &      0.50 & 08 23 59.5 & -36 37 00 &
94 &       8.0 & $        -115$ $\pm$ 13 \hspace{0.5mm} \\
256.14 &      0.25 & 08 25 11.9 & -37 24 00 &
92 &       5.3 & $+44$ $\pm$ 17 \hspace{0.5mm} \\
256.64 &     -0.22 & 08 24 43.7 & -38 04 40 &
104 &       4.7 & $+172$ $\pm$ 15 \hspace{0.5mm} \\
257.47 &      0.54 & 08 30 20.4 & -38 18 27 &
46 &       9.5 & $+23$ $\pm$ 20 \hspace{0.5mm} \\
257.71 &     -0.66 & 08 26 02.2 & -39 12 13 &
117 &       4.0 & $+144$ $\pm$ 20 \hspace{0.5mm} \\
257.92 &      0.65 & 08 32 07.7 & -38 36 33 &
383 &       2.3 & $+76$ $\pm$ 14 \hspace{0.5mm} \\
258.52 &       1.02 & 08 35 30.1 & -38 52 23 &
118 &       2.7 & $+196$ $\pm$ 26 \hspace{0.5mm} \\
258.77 &     0.08 & 08 32 23.4 & -39 37 38 &
131 &       3.4 & $+221$ $\pm$ 16 \hspace{0.5mm} \\
259.05 &     -0.72 & 08 29 49.9 & -40 19 35 &
123 &       5.7 & $+175$ $\pm$ 12 \hspace{0.5mm} \\
259.05 &     -0.75 & 08 29 45.1 & -40 20 52 &
103 &       10.0 & $+150$ $\pm$ 12 \hspace{0.5mm} \\
259.77 &       1.22 & 08 40 15.2 & -39 44 39 &
33 &       5.8 & $+250$ $\pm$ 29 \hspace{0.5mm} \\
260.41 &     -0.43 & 08 35 21.8 & -41 14 50 &
101 &       3.0 & $+221$ $\pm$ 18 \hspace{0.5mm} \\
260.52 &     -0.55 & 08 35 11.4 & -41 24 44 &
25 &       5.8 & $+247$ $\pm$ 31 \hspace{0.5mm} \\
260.69 &     -0.23 & 08 37 05.4 & -41 21 08 &
224 &       3.2 & $+204$ $\pm$ 12 \hspace{0.5mm} \\
263.20 &       1.07 & 08 50 56.2 & -42 30 54 &
165 &       5.6 & $+739$ $\pm$ 14 \hspace{0.5mm} \\
263.22 &       1.08 & 08 51 02.6 & -42 31 32 &
164 &       4.4 & $+826$ $\pm$ 19 \hspace{0.5mm} \\
263.50 &      0.17 & 08 48 12.7 & -43 19 14 &
139 &      0.9 & $+260$ $\pm$ 28 \hspace{0.5mm} \\
264.24 &      0.88 & 08 53 48.3 & -43 25 59 &
112 &       11.0 & $+406$ $\pm$ \hspace*{1.8mm}9 \hspace{0.5mm} \\
265.69 &      0.85 & 08 58 54.5 & -44 33 41 &
106 &       1.8 & $+70$ $\pm$ 28 \hspace{0.5mm} \\
266.14 &       1.08 & 09 01 33.6 & -44 44 49 &
66 &       3.2 & $+211$ $\pm$ 29 \hspace{0.5mm} \\
266.27 &      0.66 & 09 00 15.2 & -45 07 17 &
95 &       3.4 & $+396$ $\pm$ 18 \hspace{0.5mm} \\
267.03 &     0.04 & 09 00 27.1 & -46 06 15 &
58 &       3.8 & $+298$ $\pm$ 21 \hspace{0.5mm} \\
267.17 &      0.47 & 09 02 50.7 & -45 55 36 &
598 &      0.3 & $+323$ $\pm$ 28 \hspace{0.5mm} \\
268.62 &      0.58 & 09 08 56.5 & -46 55 53 &
231 &       2.1 & $+256$ $\pm$ 15 \hspace{0.5mm} \\
269.05 &      0.17 & 09 08 54.5 & -47 31 15 &
118 &       3.6 & $+456$ $\pm$ 16 \hspace{0.5mm} \\
269.55 &      0.45 & 09 12 09.7 & -47 41 35 &
156 &       1.9 & $+137$ $\pm$ 18 \hspace{0.5mm} \\
270.56 &     -0.85 & 09 10 32.9 & -49 19 14 &
58 &       5.5 & $        -152$ $\pm$ 23 \hspace{0.5mm} \\
270.91 &      0.93 & 09 19 52.0 & -48 20 10 &
57 &       6.2 & $        -149$ $\pm$ 22 \hspace{0.5mm} \\
271.22 &     -0.35 & 09 15 37.0 & -49 27 04 &
66 &       2.5 & $+215$ $\pm$ 29 \hspace{0.5mm} \\
271.30 &    -0.06 & 09 17 12.3 & -49 18 17 &
83 &       5.0 & $+136$ $\pm$ 16 \hspace{0.5mm} \\
271.52 &      -1.01 & 09 13 54.3 & -50 07 34 &
120 &       1.7 & $+75$ $\pm$ 28 \hspace{0.5mm} \\
271.70 &     -0.38 & 09 17 30.7 & -49 49 16 &
116 &       3.5 & $         -93$ $\pm$ 18 \hspace{0.5mm} \\
272.36 &      0.62 & 09 24 45.8 & -49 34 44 &
21 &       5.6 & $        -296$ $\pm$ 28 \hspace{0.5mm} \\
273.46 &      0.68 & 09 29 58.1 & -50 17 41 &
18 &       13.6 & $        -106$ $\pm$ 19 \hspace{0.5mm} \\
273.57 &       1.28 & 09 33 00.8 & -49 55 47 &
229 &       1.9 & $+20$ $\pm$ 25 \hspace{0.5mm} \\
274.77 &      0.25 & 09 34 14.3 & -51 30 18 &
51 &       3.9 & $         -76$ $\pm$ 22 \hspace{0.5mm} \\
275.02 &      0.82 & 09 37 51.7 & -51 15 07 &
26 &       8.4 & $        -101$ $\pm$ 25 \hspace{0.5mm} \\
275.48 &     -0.68 & 09 33 32.0 & -52 40 18 &
282 &       2.8 & $+248$ $\pm$ 13 \hspace{0.5mm} \\
275.56 &       1.02 & 09 41 17.5 & -51 27 26 &
281 &       6.3 & $+16$ $\pm$ \hspace*{1.8mm}7 \hspace{0.5mm} \\
275.56 &     -0.20 & 09 36 03.4 & -52 22 03 &
46 &       2.9 & $+112$ $\pm$ 30 \hspace{0.5mm} \\
275.83 &      0.16 & 09 38 55.3 & -52 16 41 &
170 &       9.3 & $        -107$ $\pm$ \hspace*{1.8mm}7 \hspace{0.5mm} \\
275.86 &      0.94 & 09 42 25.0 & -51 42 48 &
29 &       11.0 & $           0$ $\pm$ 21 \hspace{0.5mm} \\
276.46 &      0.89 & 09 45 09.7 & -52 08 29 &
153 &       2.5 & $         -17$ $\pm$ 20 \hspace{0.5mm} \\
277.44 &      0.86 & 09 49 58.5 & -52 47 04 &
10 &       12.1 & $         -71$ $\pm$ 28 \hspace{0.5mm} \\
277.78 &     -0.73 & 09 44 52.1 & -54 13 45 &
127 &       2.0 & $         -36$ $\pm$ 23 \hspace{0.5mm} \\
277.78 &     -0.81 & 09 44 32.3 & -54 17 32 &
690 &       7.3 & $         -77$ $\pm$ \hspace*{1.8mm}3 \hspace{0.5mm} \\
278.04 &      0.75 & 09 52 37.5 & -53 15 08 &
635 &       7.8 & $        -104$ $\pm$ \hspace*{1.8mm}3 \hspace{0.5mm} \\
278.37 &      0.15 & 09 51 47.9 & -53 55 46 &
18 &       13.7 & $          -4$ $\pm$ 24 \hspace{0.5mm} \\
278.43 &      0.53 & 09 53 42.8 & -53 40 12 &
42 &       4.4 & $+20$ $\pm$ 27 \hspace{0.5mm} \\
278.47 &     -0.30 & 09 50 22.2 & -54 20 21 &
22 &       6.6 & $        -116$ $\pm$ 24 \hspace{0.5mm} \\
279.04 &     -0.88 & 09 50 55.5 & -55 09 04 &
242 &       2.4 & $+239$ $\pm$ 17 \hspace{0.5mm} \\
279.09 &      0.89 & 09 58 46.2 & -53 47 21 &
47 &       5.0 & $        -120$ $\pm$ 25 \hspace{0.5mm} \\
279.15 &     -0.63 & 09 52 36.5 & -55 01 08 &
36 &       11.2 & $+341$ $\pm$ 24 \hspace{0.5mm} \\
279.15 &     -0.65 & 09 52 32.6 & -55 02 37 &
44 &       6.2 & $+329$ $\pm$ 20 \hspace{0.5mm} \\
279.33 &      0.80 & 09 59 41.3 & -54 00 11 &
35 &       7.8 & $        -157$ $\pm$ 18 \hspace{0.5mm} \\
279.80 &       1.21 & 10 03 53.7 & -53 57 24 &
25 &       9.7 & $        -145$ $\pm$ 25 \hspace{0.5mm} \\
280.53 &      0.81 & 10 06 18.9 & -54 42 57 &
37 &       5.4 & $         -10$ $\pm$ 23 \hspace{0.5mm} \\
280.62 &     -0.14 & 10 02 53.8 & -55 32 09 &
134 &       3.7 & $        -112$ $\pm$ 15 \hspace{0.5mm} \\
282.07 &     -0.78 & 10 08 36.2 & -56 54 09 &
723 &      0.9 & $+862$ $\pm$ 16 \hspace{0.5mm} \\
282.46 &      0.24 & 10 15 09.3 & -56 17 28 &
109 &       2.0 & $+256$ $\pm$ 23 \hspace{0.5mm} \\
284.30 &      0.81 & 10 28 39.3 & -56 48 09 &
80 &       2.4 & $        -547$ $\pm$ 25 \hspace{0.5mm} \\
285.15 &      0.96 & 10 34 33.2 & -57 06 17 &
268 &      0.9 & $+168$ $\pm$ 31 \hspace{0.5mm} \\
285.60 &      0.62 & 10 36 12.7 & -57 37 38 &
44 &       3.3 & $+368$ $\pm$ 35 \hspace{0.5mm} \\
286.04 &      -1.05 & 10 32 43.2 & -59 17 54 &
2826 &      0.4 & $+809$ $\pm$ 14 \hspace{0.5mm} \\
286.89 &      0.59 & 10 44 37.5 & -58 16 43 &
58 &       3.0 & $+324$ $\pm$ 28 \hspace{0.5mm} \\
288.27 &     -0.70 & 10 49 34.1 & -60 03 21 &
255 &      0.6 & $+491$ $\pm$ 25 \hspace{0.5mm} \\
290.81 &      0.74 & 11 12 45.4 & -59 47 27 &
1858 &      0.4 & $+419$ $\pm$ 23 \hspace{0.5mm} \\
292.90 &    -0.02 & 11 26 31.6 & -61 13 37 &
52 &       6.6 & $+349$ $\pm$ 28 \hspace{0.5mm} \\
293.39 &      0.73 & 11 32 19.4 & -60 40 22 &
67 &       5.0 & $+121$ $\pm$ 20 \hspace{0.5mm} \\
293.73 &      0.63 & 11 34 43.2 & -60 52 06 &
30 &       3.4 & $+116$ $\pm$ 23 \hspace{0.5mm} \\
294.29 &     -0.90 & 11 35 31.6 & -62 29 13 &
32 &       4.2 & $+449$ $\pm$ 26 \hspace{0.5mm} \\
294.38 &     -0.75 & 11 36 37.4 & -62 22 07 &
265 &      0.7 & $+470$ $\pm$ 24 \hspace{0.5mm} \\
295.17 &     0.01 & 11 44 55.1 & -61 51 26 &
57 &       3.5 & $+363$ $\pm$ 22 \hspace{0.5mm} \\
295.23 &      -1.05 & 11 43 02.3 & -62 53 56 &
114 &       2.1 & $        -207$ $\pm$ 22 \hspace{0.5mm} \\
295.29 &      -1.23 & 11 43 05.7 & -63 04 55 &
226 &       1.6 & $         -43$ $\pm$ 29 \hspace{0.5mm} \\
296.18 &     -0.59 & 11 52 05.2 & -62 40 59 &
118 &       3.7 & $+752$ $\pm$ 14 \hspace{0.5mm} \\
296.90 &      0.14 & 11 59 31.2 & -62 07 15 &
452 &       1.6 & $+1113$ $\pm$ 11 \hspace{0.5mm} \\
297.67 &      0.77 & 12 06 52.0 & -61 38 54 &
283 &       2.6 & $+570$ $\pm$ 11 \hspace{0.5mm} \\
299.42 &     -0.23 & 12 20 29.8 & -62 53 37 &
55 &       4.0 & $+535$ $\pm$ 30 \hspace{0.5mm} \\
299.51 &      -1.10 & 12 20 24.8 & -63 46 10 &
192 &       2.9 & $+315$ $\pm$ 21 \hspace{0.5mm} \\
300.25 &    -0.01 & 12 27 57.6 & -62 45 42 &
109 &       4.1 & $+123$ $\pm$ 14 \hspace{0.5mm} \\
300.47 &     -0.99 & 12 29 06.5 & -63 45 08 &
112 &       4.9 & $+412$ $\pm$ 18 \hspace{0.5mm} \\
300.65 &     -0.41 & 12 31 12.4 & -63 11 40 &
809 &      0.9 & $+358$ $\pm$ 19 \hspace{0.5mm} \\
301.14 &    -0.09 & 12 35 39.6 & -62 54 30 &
375 &       1.3 & $+350$ $\pm$ 18 \hspace{0.5mm} \\
301.70 &      0.25 & 12 40 45.4 & -62 36 01 &
21 &       5.2 & $+296$ $\pm$ 33 \hspace{0.5mm} \\
302.60 &      -1.17 & 12 48 26.2 & -64 02 26 &
162 &       1.7 & $+159$ $\pm$ 25 \hspace{0.5mm} \\
303.30 &      0.51 & 12 54 36.3 & -62 21 25 &
149 &       3.5 & $        -370$ $\pm$ 14 \hspace{0.5mm} \\
304.53 &       1.00 & 13 05 00.1 & -61 49 31 &
32 &       6.1 & $+40$ $\pm$ 28 \hspace{0.5mm} \\
305.62 &      -1.16 & 13 15 52.5 & -63 54 09 &
44 &       10.5 & $         -61$ $\pm$ 28 \hspace{0.5mm} \\
306.87 &     0.02 & 13 25 47.5 & -62 35 15 &
215 &       1.0 & $        -197$ $\pm$ 27 \hspace{0.5mm} \\
306.92 &     -0.70 & 13 27 00.7 & -63 17 46 &
275 &       2.0 & $+52$ $\pm$ 16 \hspace{0.5mm} \\
307.20 &     -0.84 & 13 29 39.3 & -63 23 47 &
41 &       8.1 & $+382$ $\pm$ 18 \hspace{0.5mm} \\
308.64 &     -0.62 & 13 41 56.1 & -62 55 38 &
40 &       2.8 & $        -133$ $\pm$ 27 \hspace{0.5mm} \\
308.73 &     0.07 & 13 41 33.5 & -62 14 06 &
92 &       1.4 & $        -661$ $\pm$ 29 \hspace{0.5mm} \\
308.93 &      0.40 & 13 42 41.5 & -61 52 38 &
152 &       2.5 & $        -752$ $\pm$ 17 \hspace{0.5mm} \\
309.06 &      0.84 & 13 43 00.9 & -61 24 52 &
670 &       2.3 & $        -504$ $\pm$ 10 \hspace{0.5mm} \\
310.20 &      -1.04 & 13 56 09.7 & -62 59 30 &
46 &       8.5 & $        -584$ $\pm$ 19 \hspace{0.5mm} \\
312.37 &    -0.04 & 14 11 37.1 & -61 25 54 &
129 &       2.3 & $        -438$ $\pm$ 28 \hspace{0.5mm} \\
313.96 &     -0.76 & 14 26 14.9 & -61 34 58 &
88 &       2.3 & $        -480$ $\pm$ 22 \hspace{0.5mm} \\
313.99 &      0.94 & 14 21 36.9 & -59 59 01 &
507 &       1.8 & $        -828$ $\pm$ 17 \hspace{0.5mm} \\
314.02 &       1.01 & 14 21 40.3 & -59 54 20 &
757 &       1.7 & $        -579$ $\pm$ 20 \hspace{0.5mm} \\
314.50 &      0.30 & 14 27 14.8 & -60 24 07 &
86 &       5.3 & $        -738$ $\pm$ 19 \hspace{0.5mm} \\
314.82 &      0.89 & 14 27 57.3 & -59 43 58 &
92 &       3.7 & $        -507$ $\pm$ 25 \hspace{0.5mm} \\
316.64 &       1.15 & 14 40 15.0 & -58 47 30 &
64 &       4.6 & $        -525$ $\pm$ 27 \hspace{0.5mm} \\
317.54 &     -0.57 & 14 52 27.0 & -59 58 11 &
152 &       2.1 & $+395$ $\pm$ 27 \hspace{0.5mm} \\
318.53 &      0.30 & 14 56 18.1 & -58 44 29 &
326 &       1.0 & $+53$ $\pm$ 29 \hspace{0.5mm} \\
319.34 &       1.08 & 14 58 58.6 & -57 40 31 &
1101 &      0.7 & $+241$ $\pm$ 21 \hspace{0.5mm} \\
319.39 &      0.74 & 15 00 28.7 & -57 57 04 &
85 &       1.4 & $+279$ $\pm$ 30 \hspace{0.5mm} \\
320.83 &      0.88 & 15 09 19.7 & -57 07 19 &
454 &      0.8 & $          -8$ $\pm$ 20 \hspace{0.5mm} \\
321.48 &       1.02 & 15 12 54.7 & -56 40 20 &
933 &       2.0 & $        -243$ $\pm$ \hspace*{1.8mm}8 \hspace{0.5mm} \\
321.58 &     -0.76 & 15 20 29.3 & -58 07 41 &
1831 &       1.1 & $        -138$ $\pm$ \hspace*{1.8mm}9 \hspace{0.5mm} \\
322.05 &     -0.95 & 15 24 17.5 & -58 01 49 &
78 &       5.4 & $        -397$ $\pm$ 17 \hspace{0.5mm} \\
323.15 &     -0.52 & 15 29 19.5 & -57 03 49 &
47 &       2.9 & $+83$ $\pm$ 25 \hspace{0.5mm} \\
324.77 &      0.61 & 15 34 10.5 & -55 12 30 &
161 &       3.9 & $         -66$ $\pm$ 14 \hspace{0.5mm} \\
325.81 &       1.08 & 15 38 05.1 & -54 13 08 &
608 &      0.6 & $         -15$ $\pm$ 31$^{\;d}$ \\
325.83 &     -0.30 & 15 43 57.7 & -55 18 26 &
247 &       1.8 & $+356$ $\pm$ 18 \hspace{0.5mm} \\
326.69 &      -1.16 & 15 52 28.2 & -55 27 04 &
95 &       9.3 & $        -142$ $\pm$ 22 \hspace{0.5mm} \\
327.31 &      0.88 & 15 47 01.4 & -53 28 29 &
256 &       1.2 & $        -189$ $\pm$ 26$^{\;d}$ \\
328.36 &     -0.41 & 15 58 00.3 & -53 48 30 &
493 &      0.3 & $        -721$ $\pm$ 35 \hspace{0.5mm} \\
329.48 &      0.22 & 16 00 57.1 & -52 36 04 &
599 &      0.7 & $        -100$ $\pm$ 25 \hspace{0.5mm} \\
330.12 &      -1.08 & 16 09 49.5 & -53 08 33 &
218 &       3.0 & $        -931$ $\pm$ 25 \hspace{0.5mm} \\
332.14 &       1.03 & 16 10 10.8 & -50 13 21 &
151 &       2.8 & $        -754$ $\pm$ 22$^{\;d}$ \\
333.72 &     -0.27 & 16 22 55.5 & -50 03 31 &
108 &       2.5 & $+204$ $\pm$ 29 \hspace{0.5mm} \\
335.32 &      0.60 & 16 26 00.4 & -48 18 36 &
352 &       2.9 & $        -138$ $\pm$ 11 \hspace{0.5mm} \\
337.06 &      0.85 & 16 32 01.7 & -46 52 39 &
100 &       2.5 & $        -739$ $\pm$ 32 \hspace{0.5mm} \\
339.65 &     -0.24 & 16 46 44.6 & -45 40 00 &
222 &       2.2 & $        -398$ $\pm$ 19 \hspace{0.5mm} \\
342.16 &     -0.74 & 16 57 52.7 & -44 02 49 &
264 &       2.2 & $+127$ $\pm$ 15 \hspace{0.5mm} \\
342.62 &     -0.45 & 16 58 15.4 & -43 30 19 &
50 &       3.9 & $        -913$ $\pm$ 30 \hspace{0.5mm} \\
343.29 &      0.60 & 16 56 01.8 & -42 19 58 &
102 &       4.7 & $       -1035$ $\pm$ 16 \hspace{0.5mm} \\
345.22 &      0.68 & 17 02 02.9 & -40 45 50 &
39 &       12.6 & $+183$ $\pm$ 21 \hspace{0.5mm} \\
347.40 &      -1.04 & 17 16 09.8 & -40 02 24 &
72 &       4.0 & $        -524$ $\pm$ 36 \hspace{0.5mm} \\
349.65 &     -0.36 & 17 19 58.3 & -37 48 15 &
133 &       3.3 & $+110$ $\pm$ 20 \hspace{0.5mm} \\
350.52 &     -0.73 & 17 23 58.9 & -37 18 18 &
127 &       2.2 & $+277$ $\pm$ 32 \hspace{0.5mm} \\
351.31 &     -0.53 & 17 25 22.9 & -36 32 16 &
200 &       3.3 & $        -247$ $\pm$ 15 \hspace{0.5mm} \\
351.82 &      0.17 & 17 23 56.8 & -35 43 34 &
118 &       6.1 & $+134$ $\pm$ 16 \hspace{0.5mm} \\
352.13 &       1.15 & 17 20 51.7 & -34 54 49 &
116 &       6.5 & $+76$ $\pm$ 30 \hspace{0.5mm} \\
355.43 &     -0.81 & 17 37 32.3 & -33 14 40 &
76 &       4.2 & $+601$ $\pm$ 21 \hspace{0.5mm} \\
356.57 &      0.87 & 17 33 45.9 & -31 22 37 &
116 &       1.6 & $+985$ $\pm$ 30 \hspace{0.5mm} \\
 \enddata
\tablenotetext{a}{The identified location is the peak of the gaussian
fit to the source in polarised intensity.  All sources were either
unresolved or partially resolved.}
\tablenotetext{b}{$\;I$ is the peak-pixel Stokes I  value of the
    interferometric data.}
\tablenotetext{c}{$\;m$ is the fractional polarisation (linear polarised
intensity divided by Stokes I) averaged over the FWHM pixels used for RM calculation.}
\tablenotetext{d}{A RM for this source was calculated by \citet{bmg01}
using a simpler approach
than the more rigorous method used here. These new values should replace
the previously determined values.}
\label{tab1}
\end{deluxetable}
%

\clearpage

\clearpage

\clearpage

\clearpage

\clearpage

\end{document}